\documentclass[conference,lettersize]{IEEEtran}
\usepackage{amsmath,amsfonts}
\usepackage{array}
\usepackage[caption=false,font=normalsize,labelfont=sf,textfont=sf]{subfig}
\usepackage{algorithmic}
\usepackage{algorithm}
\usepackage{multirow}
\usepackage{textcomp}
\usepackage{stfloats}
\usepackage{url}
\usepackage{tabularx}
\usepackage{verbatim}
\usepackage{dsfont}
\usepackage{bm}
\usepackage{graphicx}
\usepackage{tabularray}
\usepackage[table,xcdraw,dvipsnames]{xcolor}
\usepackage{amsmath}
\usepackage{hyperref}
\usepackage{cite}

\usepackage{enumitem}

\algsetup{ linenosize=\small, linenodelimiter=. }
\usepackage{tikz}
\usetikzlibrary{matrix, positioning, patterns, shapes, arrows}
\usepackage{pgfplots}
 
\usepgfplotslibrary{groupplots}
\usepgfplotslibrary{colorbrewer}
\usetikzlibrary{spy,backgrounds}
\usepackage{epsfig}
\usepackage{pgfplots}
\pgfplotsset{compat=newest}
\usepgfplotslibrary{groupplots}
\usepackage{tikz}
\usetikzlibrary{matrix, positioning, patterns, shapes, arrows}
\usepgfplotslibrary{colorbrewer}
\pgfplotsset{compat = 1.15, cycle list/Set1-8} 
\usetikzlibrary{pgfplots.statistics, pgfplots.colorbrewer} 

\usepackage{pgfplotstable}

\definecolor{Paired-1}{RGB}{31,120,180}
\definecolor{Paired-2}{RGB}{166,206,227}
\definecolor{Paired-3}{RGB}{51,160,44}
\definecolor{Paired-4}{RGB}{178,223,138}
\definecolor{Paired-5}{RGB}{227,26,28}
\definecolor{Paired-6}{RGB}{251,154,153}
\definecolor{Paired-7}{RGB}{255,127,0}
\definecolor{Paired-8}{RGB}{253,191,111}
\definecolor{Paired-9}{RGB}{106,61,154}
\definecolor{Paired-10}{RGB}{202,178,214}
\definecolor{Paired-11}{RGB}{177,89,40}
\definecolor{Paired-12}{RGB}{105,105,105}
\definecolor{Paired-13}{RGB}{80,80,80}

\definecolor{Color1}{HTML}{D81B60}
\definecolor{Color2}{HTML}{1E88E5}
\definecolor{Color3}{HTML}{FFC107}
\definecolor{Color4}{HTML}{004D40}
\newcommand*\idelta{$I_\Delta$ }
\newcommand*\pcfrozen{$PC_{Frozen}$ }
\begin{document}

\title{Hardware-friendly IR-HARQ for Polar SCL Decoders}
\author{\IEEEauthorblockN{Marwan Jalaleddine, Jiajie Li, Warren J. Gross} 
\IEEEauthorblockA{\textit{Department of Electrical and Computer Engineering, McGill University,} 
Montreal, Quebec, Canada \\
Emails: marwan.jalaleddine@mail.mcgill.ca, jiajie.li@mail.mcgill.ca,   warren.gross@mcgill.ca}}


\maketitle

\begin{abstract}
To extend the applications of polar codes within next-generation wireless communication systems, it is essential to incorporate support for Incremental Redundancy (IR) Hybrid Automatic Repeat Request (HARQ) schemes. The baseline IR-HARQ scheme's reliance on set-based operations leads to irregular memory access patterns, posing significant challenges for efficient hardware implementation. Furthermore, the introduction of new bit types increases the number of fast nodes that are decoded without traversing the sub-tree, resulting in a substantial area overhead when implemented in hardware. To address these issues and improve hardware compatibility, we propose transforming the set-based operations within the polar IR-HARQ scheme into binary vector operations. Additionally, we introduce a new fast node integration approach that avoids increasing the number of fast nodes, thereby minimizing the associated area overhead. Our proposed scheme results in a memory overhead of $25-27\%$ compared to successive cancellation list (SCL) decoding without IR-HARQ support.
\end{abstract}


\section{Introduction}
Polar codes are a capacity-achieving error-correcting code \cite{arikan_channel_2009}. Due to their low encoding and decoding complexity and their competitive error-correcting performance with SCL decoding \cite{tal_list_2015}, polar codes have been implemented in modern communication systems, such as the 5G New Radio (NR) standard \cite{3gpp_nr_2021}, particularly in the control channels. To enhance their applicability, especially for the data channel, support for HARQ, which aims to enhance link reliability through re-transmissions, is necessary.

HARQ schemes are typically categorized into two main schemes: Chase combining (CC) and incremental redundancy. In CC-HARQ, the same coded block is re-transmitted, offering simplicity but limiting improvements in both throughput and reliability \cite{cheng_coding_2006}. Conversely, IR-HARQ introduces additional redundancy bits with each re-transmission, which improves performance \cite{cheng_coding_2006}, but efficient hardware-compatible implementations of IR-HARQ for polar codes are yet to be developed. 


CC-HARQ approaches were first proposed in  \cite{chen_hybrid_2013,chen_polar_2014,el-khamy_harq_2015,tavildar_h-arq_2016} for polar codes by selectively puncturing and re-transmitting bits for which the selection is determined through greedy search algorithms.
More recently, a puncturing pattern permutation scheme was proposed 
 \cite{zhang_hybrid_2018}, in which the puncturing pattern used in the initial transmission is permuted to generate the puncturing patterns for HARQ re-transmissions.  Additionally, \cite{chen2022equivalent} determines the codeword bits for puncturing by flipping the binary representations of the indices of the initially punctured bits. However, for all of these methods, the puncturing patterns must be carefully designed, as they significantly impact the capacity of the virtual sub-channels allocated for message bits. Additionally,  these approaches suffer from significant performance losses in error correction, attributed to the Chase-combining of log-likelihood ratios (LLRs) \cite{cheng_coding_2006,zhao_adaptive_2018}. 



To reduce the degradation in decoding performance achieved with CC-HARQ schemes, \cite{li_capacity-achieving_2016} proposed a polar IR-HARQ scheme featuring an incremental freezing mechanism in which unreliable information bits from previous transmissions are retransmitted, decoded, and progressively frozen to facilitate the decoding of the initial transmission. However, as the two transmissions are decoded independently as short codes, this approach is unable to achieve the coding gains attributed to longer codes.

This method was subsequently improved by \cite{zhao_adaptive_2018}, which proposed concatenating multiple transmissions into a longer polar code, thus enhancing both coding and diversity gains. This improvement was achieved by incrementally extending the polar encoding matrix, leveraging the superior decoding performance associated with longer polar codes \cite{arikan_channel_2009}. 
Despite the advancements made by \cite{zhao_adaptive_2018}, their proposed algorithm has yet to be adapted into a hardware-friendly form or implemented in hardware. This limitation arises first from the algorithm's dependence on set-based operations, which cause irregular memory accesses. Second, the newly introduced bit types restrict the algorithm's use of  \textit{fast nodes}, which are nodes that are decoded without traversing the sub-tree. To integrate fast nodes into polar IR-HARQ and to reduce the latency of successive cancellation decoding by limiting how deeply we traverse the decoding graph, it is essential to implement all possible fast node types; however, the set of possible fast nodes is significantly large due to the introduction of new bit types in the IR-HARQ scheme. For example, in a node size of 4, the number of fast nodes increases from 16 in the SCL decoder to 81 in the naive IR-HARQ implementation. This naive approach would result in considerable area overhead in hardware.

This work proposes simplifications and hardware-compatible operations to enable the implementation of polar IR-HARQ. By representing all possible fast nodes as linear combinations of the outputs of the basic fast node types and the values of a newly introduced bit type, the number of fast nodes required for hardware implementation stays the same as it was with SCL with no IR-HARQ support. Additionally, by describing the set-based operations in terms of binary vector-based operations, the irregular memory accesses issue is avoided. The structure of this work is as follows: first, a review of polar codes and the polar IR-HARQ scheme is provided. Next, a hardware-friendly algorithm is detailed and impacts on memory, the number of computations, and latency are discussed. Finally, the conclusions are presented.

\section{Preliminaries}
\subsection{Notation}
Matrices are denoted by a bold upper-case letter ($\bm{M}$), while vectors are denoted by bold lower-case letters ($\bm{v}$). The $i^{\text{th}}$ element of a vector $\bm{v}$ is denoted as $\bm{v}_i$. The sub-vector obtained by taking a slice from index $i$ to $j$ of vector $\bm{v}$ is denoted as $\bm{v}_{i:j}$. All indices start at $1$. In this work, all operations are restricted to the Galois field with 2 elements, noted $\mathbb{F}_2$. Operations $\lor$, $\land$, $\oplus$ represent OR, AND and XOR binary bit-wise operations respectively. $|\cdot|$ denotes set size, and $\mathds{1}$ denotes a vector with all ones.  Furthermore, this work considers $(n,k)$ linear block codes, where $n$ is the code length and $k$ is the code dimension.
\subsection{Polar Codes}
\label{sec:PC}

A polar code with a mother code length $n = 2^m$ is constructed by combining two polar codes of length $\frac{n}{2}$. Let us consider an input set $\mathbf{u}$ and a coded set $\mathbf{x}$ each of length $n$. The recursive encoding process can be expressed as a modulo-2 matrix multiplication as in
\begin{equation}
\mathbf{x} = \mathbf{u} \cdot \mathbf{G}^{\otimes m}\text{,} \label{eq1}
\end{equation}
where $\mathbf{G}^{\otimes m}$ is the $m$-th Kronecker product of the polarizing matrix $\mathbf{G}=\bigl[\begin{smallmatrix} 1&0\\ 1&1 \end{smallmatrix} \bigr]$. We will refer to this encoding process as ascending the tree since it can be done using by ascending polar encoding tree where each stage $s$ in the tree is a polar code of length $2^s$ \cite{arikan_channel_2009}. We also refer to each polar sub-code in a polar tree as a node.

The process of polar encoding involves determining the $k$ most reliable bit-channels after puncturing the polar code. The resulting length of the polar code after puncturing is denoted by $N$. Frozen bits are the $N-k$ least reliable non-punctured bits that are set to a predefined value (typically $0$) which is known to the decoder. The polar decoder receives $\mathbf{y}$ as its noisy channel output using which, the log likelihood ratio (LLR) can be calculated. 

Successive Cancellation (SC) decoding provides each bit estimate $\hat{u}_i$ based on $\mathbf{y}$, the previously estimated bits $\mathbf{\hat{u}}$, and the location of frozen bits $\mathcal{F}$. The LLR-based formulation is
\begin{equation}
\hat{u}_i =
  \begin{cases}
    0, & \text{if } i \in \mathcal{F} \lor \log\frac{\text{P}(\mathbf{y}_i,\mathbf{\hat{u}}_i|\hat{\bm{u}}_i = 0)}{\text{P}(\mathbf{y}_i,\mathbf{\hat{\bm{u}}}_i|\hat{u}_i = 1)}\geq 0,\\
    1, & \text{otherwise}.
  \end{cases} \label{eq2}
\end{equation}

Two types of messages are passed through the different levels in a decoder tree: 
 \begin{enumerate}
     \item the soft messages which contain the LLR values $\alpha$ at a certain node calculated as \cite{leroux_semi-parallel_2013}:
     \begin{align}
\bm{\alpha}^{left}_i =& \text{sgn}(\bm{\alpha}_{i})\text{sgn}(\bm{\alpha}_{i+2^{s-1}})\min(|\bm{\alpha}_i|,|\bm{\alpha}_{i+2^{s-1}}|)  \text{,} \nonumber\\
\bm{\alpha}^{right}_i =& 
 \bm{\alpha}_{i+2^{s-1}} + (1-2\bm{\beta}^{left}_i)\bm{\alpha}_i \text{,}
\label{eq3}
\end{align}
     \item the hard bit estimates $\bm{\beta}$. Each node at level $s$ of the decoder tree contains $2^s$ bits and the messages are calculated as \cite{leroux_semi-parallel_2013}:
     \begin{equation}
\bm{\beta}_i =
  \begin{cases}
    \bm{\beta}^{left}_i \oplus \bm{\beta}_i^{right}, & \text{if} \quad i \leq 2^{s-1}\\
    \bm{\beta}_{i-2^s}^{right}, & \text{otherwise},
  \end{cases}
  \label{eq4}
\end{equation}
where $\bm{\beta}_i$ are called \emph{partial sums} and $\text{sgn(x)}$ produces the sign of $x$. We will will refer to calculating $\bm{\beta}^{left}$ and $\bm{\beta}^{right}$ as descending the tree.
 \end{enumerate} 

 Instead of traversing the entire tree down to stage $0$, special sub-trees/nodes of the polar decoding tree can be decoded efficiently if specific patterns of the frozen and information bits are encountered \cite{sarkis_fast_2014}. We denote the number of bits and corresponding leaf nodes in the sub-tree as $N_{v}$.
 To name a few, the rate $1$ node has all its leaf nodes being information bits, rate 0 node has all its leaf nodes being frozen bits, the repetition node has its last  bit being an information bit while all other bits are frozen bits, and the single parity check node has its first leaf node being frozen bits while all other bits are information bits. 

Successive Cancellation List (SCL) decoding generates two tentative paths, corresponding to $\bm{\hat{u}}_i = 0$ and $\bm{\hat{u}}_i = 1$, on the decoding tree for each non-frozen bit in order to enhance the error-correcting performance of SC decoding. Only the $L$ most-likely paths are retained in order to prevent an exponential increase in the number of tentative paths. Specifically, in LLR-based SCL decoding~\cite{balatsoukas-stimming_hardware_2014}, the $L$ best paths are determined by using the following path metric	
\begin{equation}
\text{PM}_i^l =
  \begin{cases}
    \text{PM}_{i-1}^l, & \text{if } \hat{\bm{u}}_i^l = \frac{1}{2}\left(1-\text{sgn}\left(\bm{\alpha}_i^l\right)\right),\\
    \text{PM}_{i-1}^l + |\bm{\alpha}_i^l|, & \text{otherwise},
  \end{cases}
  \label{eq5}
\end{equation}
where $l$ is the path index and $\bm{\alpha}_i^l$ is the LLR metric at path $l$ associated with the $i$-th bit. A smaller path metric indicates a more reliable path. SCL can also leverage the use of special nodes to reduce decoding latency and complexity \cite{sarkis_fast_2016,hashemi_fast_2017}.

\subsection{Polar IR-HARQ with Matrix extension}

To begin, we define key terms for the $t$-th transmission:
\begin{itemize}
    \item $I^t$: the set of information bit channels,
    \item $\mathcal{F}^t$: the set of frozen bit channels,
    \item $PF_{\Delta}^t$: the set of frozen bit channels mapped to the value of a previously estimated bit referred to by \pcfrozen,
     \item $I_{\Delta}^t$: the set of newly added more reliable information bit channels,
     \item $RM^t$: the set of punctured bit channels,
     \item  $N^t$: the length of the punctured polar code at the $t$-th transmission.
\end{itemize}

The IR-HARQ \cite{zhao_adaptive_2018} scheme operates through the following steps:

\begin{enumerate}
    \item \textbf{Initial Setup}: 
    Determine the sets $I^1$ and $\mathcal{F}^1$ for the first ($t=1$) $N^1$ sized transmission. Set the puncturing bits $RM^1$ as the leftmost bits of the polar encoding tree. Transmit the $N^1$ coded bits over the physical channel, with $PF_{\Delta}^1$ and $I_{\Delta}^1$ initially set as empty.
    
    \item \textbf{Subsequent Transmission}: 
    For each new transmission, increment $t$ by 1. Determine the information bit ($I^*$) and frozen bit sets ($\mathcal{F}^*$) for the $N^t$-sized code with the same information block length $k$ and a new code block length $N^t$ assuming no IR-HARQ is being used.
    
    \item \textbf{Redundancy Extension}:
    Identify the newly introduced more reliable information bit channels ($I_{\Delta}$) by 
    \[
    I_{\Delta} = \Upsilon(I^* _{ N^t : N^{t-1}}, I^{t-1}_{ N^t : N^{t-1}}) \quad . 
    \]
    Here, $\Upsilon(A, B)$ represents elements in set $A$ but not in $B$. Also, compute the newly introduced \pcfrozen bit channels ($PF_{\Delta}$)  by taking the first $|I_{\Delta}|$ information bits in $I^{t-1}$ not in $I^*$:
    \[
    PF_{\Delta} = \Upsilon(I^{t-1}, I^*)_{1:|I_{\Delta}|}.
    \]

    \item \textbf{Bit Value Copying}: 
    Establish a one-to-one relationship between $PF_{\Delta}$ and $I_{\Delta}$.
    
    \item \textbf{Encoding and Transmission}: 
    Encode using the Ar{\i}kan polarizing matrix to produce code of length $N^t$. Only the difference $N^t - N^{t-1}$ coded bits from the newly extended part of the $N^t$-sized codeword are transmitted as redundant bits, as the other parts have been transmitted in previous rounds.
    
    \item \textbf{Update Sets}:
    Update the channel sets:
    \begin{align*}
        I^t &= \{I_{\Delta}, \Upsilon(I^{t-1}_{\Delta}, PF_{\Delta})\},\\
        \quad PF^t_{\Delta} &= \{PF_{\Delta}^{t-1}, PF_{\Delta}\},\\
        \quad \mathcal{F}^t &= S^t \setminus \{I^t, PF_{\Delta}^t\},
    \end{align*}
    where $S^t$ represents the index set of the $N^t$ sized code at the $t^{th}$ transmission:  $S^t=\{ 1,2,3,..., N^t\}$. 
   
\end{enumerate}

Repeat steps 2-6 until an acknowledgment is received.

\section{Hardware-friendly algorithm}
To improve hardware compatibility of the aforementioned polar IR-HARQ scheme, we transform the set-based operations, used in the polar IR-HARQ scheme, into binary vector operations. We also introduce a hardware-optimized bit-type generation method, a fast node integration scheme, and a modified control sequence for each node.
\subsection{Newly introduced bit types}
The IR-HARQ scheme introduces two new bit types, the \idelta  and the \pcfrozen bits. The \idelta bits can be considered as normal information bits; however, their bit results should be saved in memory. The \pcfrozen bits behave in a similar manner to frozen bits in SCL decoding as they do not produce extra paths; however, their values can be either $0$ or $1$ depending on the \idelta bits they are mapped to. Additionally, a node size ($N_v$) of $4$ which originally had $16$ node variations (types), corresponding to all the different combinations of information and frozen bits, would have $81$ different variations if we were to consider \pcfrozen bits as a new bit type. In this case all the 81 different node types would need to be implemented in hardware for fast list decoding \cite{sarkis_fast_2016} which would result in an added area overhead.

To avoid introducing a new bit type we consider \pcfrozen bits as a special type of frozen bits. We refer to the traditionally frozen as $Frozen\_z$ bits and we assume that their values are always fixed to 0. Assuming that we have the vectors $\mathbf{fr}$, $\mathbf{rm}$ and $\mathbf{pc}$ which indicate with a $1$ which indices are frozen, rate-matched  and \pcfrozen  bits respectively,
the information $\mathbf{iv}$ bits, the frozen bits with zero value $\mathbf{fr\_z}$ and the \idelta  bits $\mathbf{id}$ can be respectively obtained using the following equations:
\begin{align}
    \mathbf{iv}_i = &  \overline{\mathbf{fr}_i},\\
    \mathbf{fr\_z} = &  \mathbf{fr}_i \land  \overline{\mathbf{pc}_i}, \\
    \mathbf{id}_i = &  \begin{cases}  0 & i<N^1, \\
    \overline{\mathbf{fr}}_i & i>N^1.
    \end{cases}\label{eq:idelta}
\end{align}
This notation assumes that rate-matched, \pcfrozen and $Frozen\_z$ bits are considered different types of frozen bits.

\subsection{Generating bit types} \label{sec:generating-bit-types} 

Considering the IR-HARQ situation where the initial transmission of length $N^{t-1}$ is transmitted with the following frozen-bit, rate-matching and \pcfrozen sequences $
f^{t-1}= \{\mathbf{fr}^{t-1},\mathbf{rm}^{t-1},\mathbf{pc}^{t-1} \}$, and another transmission of length $p$ is requested to be transmitted, the following steps are done to generate the set of $f^{t}= \{\mathbf{fr}^{t},\mathbf{rm}^{t},\mathbf{pc}^{t} \}$:
\begin{enumerate}
    \item Generate the $\{\mathbf{fr}^{*},\mathbf{rm}^{*}\}$ for a code length $N^t$ for the code with no IR-HARQ support.
    \item Set $\mathbf{rm}^{t} =\mathbf{rm}^{*}$
    \item If the length of the mother code at the current ($n^t$) is greater than the mother code length of the previous transmission ($n^{t-1}$), extend $\mathbf{fr}^{t-1}$ and $\mathbf{rm}^{t-1}$ with 1's and $\mathbf{pc}^{t-1}$ with 0's up to a length $n^t$.

    \item Calculate \idelta as shown in (\ref{eq:idelta}) and find the \pcfrozen bits as follows:
    \begin{equation}
    \begin{split}
        \mathbf{pc}^t_i=&{}  
               \mathbf{pc}^{t-1} _i \lor \left(\vphantom{\sum_{k=0}^{N^t} \mathbf{id}_k^{t}} (\overline{\mathbf{fr}}^{t-1}_i \land \mathbf{fr}^{*}_i ) \right.\\
               & \left. \cdot \left( \sum_{j=0}^i (\mathbf{pc}^t_j \land \overline{\mathbf{pc}}^{t-1}_j)<\sum_{k=0}^{N^t} \mathbf{id}_k^{t} \right)  \right)  . 
               \end{split}
               \label{eq:pc_res}
    \end{equation}
    \item Create a lookup table $\mathbf{lut}$ mapping each new \idelta to a new \pcfrozen.
    \item  Calculate the new frozen set as follows:
    \begin{equation}
        \mathbf{fr}^t_i=
        \begin{cases}  \left(\mathbf{fr}^{t-1}_i \lor \left(\mathbf{pc}_i^t \land  \overline{\mathbf{pc}}_i^{t-1}\right)\right) 
 &, i< N^{t-1},\\
        
        \overline{\mathbf{id}}^t_i   & , otherwise.
        \end{cases}
    \end{equation}
\end{enumerate}

\subsection{Generating candidate codewords for fast nodes}
For each node, the SCL decoder first generates a pre-defined number of candidate codewords \cite{sarkis_fast_2016,hashemi_fast_2017}. Then the $L$ most-likely candidates are selected based on their path metrics.
To determine all the possible candidates at stage $s$ and with the index of the first element in the node $sp$, the node candidates at a  stage $s$ and starting index $sp$  ($\bm{\beta} (s)_{sp}$), all the different combinations of the nonfrozen $sp$ to $sp+2^{s}$ bit values at stage 0, denoted as $ \bm{\beta}(0)_{sp:sp+2^{s}}$, are encoded using the polar code generator matrix $\mathbf{G}$ :

\begin{equation}
    \bm{\beta} (s) = \bm{\beta}(0)_{sp:sp+2^{s}}.\mathbf{G}^{\otimes s} .\\
    \label{eq:Encoding_1}
\end{equation}

 Hence, given that we have only two main types of bits which are mutually exclusive ($\mathbf{iv} \oplus \mathbf{fr}=\mathds{1}$) and that $\bm{\beta}(0)_{sp:sp +2^{s}}\land \mathds{1} =\bm{\beta}(0)_{sp:sp +2^{s}}$, we can use the distributive rule to expand (\ref{eq:Encoding_1}) into:

\begin{equation}
   \begin{split}
 \bm{\beta} (s) {}&=  \left( \bm{\beta}(0)_{sp:sp +2^{s}} \land \mathbf{iv}_{sp:sp +2^{s}}\right) .\mathbf{G}^{\otimes s} , \\
     & \text{  } \oplus \left( \bm{\beta}(0)_{sp:sp +2^{s}} \land \mathbf{fr}_{sp:sp +2^{s}}\right) .\mathbf{G}^{\otimes s}   .
\end{split} \label{eq:linear_expansion_1}
\end{equation}
Using the distributive rule and noting that $\mathbf{fr\_z}_i$, $ \mathbf{pc}_{i}$ and $\mathbf{rm}_{i}$ are mutually exclusive types of $\mathbf{fr}_i$ ($\mathbf{fr\_z}_i \oplus \mathbf{pc}_{i} \oplus \mathbf{rm}_{i} =\mathbf{fr}_i$), we can expand  (\ref{eq:linear_expansion_1}) to consider all the different types of frozen bits to obtain:

\begin{equation}
    \begin{split} \bm{\beta} (s) {}&= \left( \bm{\beta}(0)_{ sp:sp+2^{s}}\land \mathbf{iv}_{ sp:sp+2^{s}}\right) .\mathbf{G}^{\otimes s} , \\
      &  \text{  } \oplus \left( \bm{\beta}(0)_{ sp:sp+2^{s}}\land \mathbf{fr\_z}_{ sp:sp+2^{s}}\right) .\mathbf{G}^{\otimes s} , \\ 
     & \text{  } \oplus \left( \bm{\beta}(0)_{ sp:sp+2^{s}}\land \mathbf{pc}_{ sp:sp+2^{s}}\right) .\mathbf{G}^{\otimes s} , \\
       & \text{  } \oplus \left( \bm{\beta}(0)_{ sp:sp+2^{s}} \land \mathbf{rm}_{ sp:sp+2^{s}}\right) .\mathbf{G}^{\otimes s} . \\
\end{split}\label{eq:linear_expansion_2}
\end{equation}
By noting that $ \bm{\beta}(0)_{ sp:sp+2^{s}}$ of $Frozen\_z$ and rate matched bits always equates to $\mathbf{0}$, (\ref{eq:linear_expansion_2}) can be simplified as:

\begin{equation}
   \begin{split}
 \bm{\beta} (s) {}&=  \left( \bm{\beta}(0)_{ sp:sp+2^{s}}\land\mathbf{iv}_{ sp:sp+2^{s}}\right) .\mathbf{G}^{\otimes s}  \\
     & \text{  } \oplus \left( \bm{\beta}(0)_{ sp:sp+2^{s}}\land\mathbf{pc}_{ sp:sp+2^{s}}\right) .\mathbf{G}^{\otimes s} . \\ \label{eq:linear_expansion_3}
\end{split}
\end{equation}

This means that the resulting candidate bits $\bm{\beta} (s)$ for a certain node can be obtained by the linear combination of the encoded information bits with the result of the encoded \pcfrozen bits. The encoding of the \pcfrozen bits can be done using a XOR array leveraging the unique structure of polar codes and all the combinations of information bits to generate the list of all the possible candidates can be generated in advance and stored in registers.
\\

\noindent \textbf{Example:}

For a repetition node of length 16, the codeword candidates of list index $l$ and bit index $i$ $(\bm{\beta_c})_{l,i}$ are all the all 1s or all 0s candidates \cite{sarkis_fast_2016}:
\begin{align}
 \bm{\beta_c} (4)_{1,:} &=  1111111111111111 ,\\
 \bm{\beta_c} (4)_{2,:} &=  0000000000000000 .
\end{align}
Assume that we encounter the $\bm{PC_{Frozen}}$ bit sequence (\ref{eq:forzenpc_ex}) at stage 0. To calculate the new candidate codewords at stage 4, we first ascend the tree to stage 4 :
\begin{align}
    \bm{PC_{Frozen}}(0) &= 0100000000000000  ,\label{eq:forzenpc_ex} \\ 
    \bm{PC_{Frozen}}(4) &= 1100000000000000 .
\end{align}
Then we can obtain the new candidate codewords $\bm{\beta'}$ through XOR-ing each of the candidates $\bm{\beta_c}$ with the \pcfrozen bits at stage 4:
 \begin{align}
 \bm{\beta'} (4)_{1,:} &=  0011111111111111, \\
 \bm{\beta'} (4)_{2,:} &=  1100000000000000.
\end{align}

\subsection{Control flow:}

\addtolength{\topmargin}{+0.13cm}
\begin{algorithm}[t]
\caption{ Control Flow for decoding a node in polar code} \label{algorithm:control-flow}
\begin{algorithmic}[1]
\REQUIRE $sp$, $s$, $L$, $\mathbf{\alpha}$, $\mathbf{\beta}$, $\mathbf{pc}$, $\mathbf{fr}$, $\mathbf{lut}$, $N_1$
\ENSURE Updated $\beta(s)$,  \bm{\pcfrozen}

\FOR {$l=1$ \TO $L$}
\STATE Encode $\mathbf{PC_{Frozen}}_{ \text{ } l, sp:sp+2^s}$ bits using the polar code generator matrix
\FOR{$i = 1$ \TO $ 2^s - 1$}
\FOR{ $j$ \TO $L_a$}
    \STATE Create new forked path $j$ from $l$
    \STATE$\beta' (s)_{l,j,i} \gets {PC_{Frozen}}_{\text{ } l, i+sp}$ \XOR $\beta_i(s) _{j,i}$
    \STATE Update PMs of new forked path based on $\mathbf{\alpha}(s)_{l,i}$ and $\beta'(s)_{l,j,i}$
\ENDFOR
\ENDFOR
\ENDFOR

\STATE Select the $L$ paths to preserve which have the minimum path metrics (PMs) from $L \times L_a$ paths 
\STATE Update  $\beta(s)_{l,i}$ based on the surviving $l$ paths' $\beta' (s)_{l,j,i}$

\FOR {$l=1$ \TO $L$}
\STATE Descend the tree from $s$ stage to stage $0$ to find the message bits 

\FOR{$i = 0$ \TO $2^s$}
    \IF{$i \geq N^1$ \AND $(\mathbf{\overline{fr}}_{i}$ $\lor$ $\mathbf{pc}_{i})$}
        \STATE $\mathbf{PC_{Frozen}}_{ \text{ } l, \mathbf{lut}_{l,sp+i}} \gets \beta(0)_{l,i}$
    \ENDIF
\ENDFOR
\ENDFOR

\end{algorithmic}
\end{algorithm}

Before receiving the channel signal, the decoder should initialize all the \pcfrozen bit values to 0. The decoding tree is then traversed through the original tree traversal method of SCL decoding until encountering a decodable node. The control flow of decoding a node with the proposed HARQ support is modified from the original control flow and is presented in Algorithm \ref{algorithm:control-flow}. There are four main added operations introduced which are: ascending the tree, calculating new candidate codewords, descending the tree and routing of $I_\Delta$ bits to \pcfrozen bits.
  The inputs to the modified node decoding algorithm are: $sp$, $s$, $L$, $\bm{\alpha}$, $\bm{\beta}$, $\mathbf{pc,fr}$, $\mathbf{lut}$ the look up table mapping \idelta to \pcfrozen, $\mathbf{pc}$ and $\mathbf{fr}$. The algorithm proceeds as follows:
\begin{itemize}[align=left]
    \item[* Line 2:] At the start of decoding a node, the \pcfrozen bits of that node and each path are encoded using the polar code generator matrix for that stage  ($\mathbf{G}^{\otimes s}$).
    \item[* Lines 3 - 9:] The \pcfrozen bits at stage (s) are XOR-ed to each of the candidate codewords for the current node to produce $L_a$ new paths from each original path. The modified candidate codewords $\bm{\beta'(s)}$ and the path metric for each of these forked paths is calculated.
    \item[* Line 11:] After all the new candidate codewords are generated, we select the paths to preserve through choosing the paths with the least path metrics.
    \item[* Line 12-14:] This is followed by updating the \pcfrozen bits for that node by first descending the tree. 
    \item[* Line 15-19:] All the non-frozen bits after the first transmission are transferred to their mapped location $\mathbf{lut}_{i}$.
    
\end{itemize}

\section{Discussion}

\subsection{Intra-node bit dependency}
The method described in \autoref{sec:generating-bit-types} permits dealing with the inter-node \pcfrozen bit dependencies without catering for the dependency happening within a certain node (intra-node dependency). For an intra-node dependency to occur, an information bit should be located to the left of a frozen bit within a single node, and both bits should be mapped to each other. Since rate-zero and rate-one nodes contain only one type of bit, intra-node dependency does not arise in these cases. Furthermore, single parity check and repetition nodes also avoid intra-node dependency because their information bits are positioned to the right of frozen bits. For other nodes, we can eliminate intra-node dependency by treating the \idelta bit as a $Frozen\_z$ bit and the \pcfrozen bit as an information bit at the encoder and decoder.
We can do that by iterating over all the indices in the node except the last index and checking whether a bit is an \idelta bit. If an \idelta bit is encountered, intra-node dependency is checked by checking if it is mapped to a \pcfrozen bit in the same node. If an intra-node dependency is encountered, the \idelta bit is converted to a frozen bit and the \pcfrozen bit is converted to an information bit.

The encoder in this scheme must be aware of how the decoder navigates the graph, which can be achieved through a one-time transmission of the minimum node size and the types of supported nodes along with their maximum size.

 \subsection{Complexity overhead} 
 The addition of HARQ imposes memory, computational and latency overheads which will be discussed in this section:
 \subsubsection{Memory Complexity}
 The memory requirement of the data and control signals( $M_{SCL}$) of SCL decoding is \cite{hashemi_fast_2019}: 
 \begin{multline}
     M_{SCL} = N  \cdot  Q_e + (N-1)  \cdot  L   \cdot Q_i  + L \cdot Q_m \\+ (2N-1) \cdot L + 2 \cdot  N.
\end{multline}
 where  $Q_e,Q_i$ and $Q_m$ are the number of bits used for the quantization of the external LLRs, internal LLRs and the path metrics respectively. The HARQ enabled SCL decoder has the following memory requirement of the data and control signals ( $M_{SCLm}$) of SCL decoding: 
\begin{multline}
    M_{SCLm} = N \cdot Q_e + (N-1) \cdot L \cdot Q_i + L \cdot Q_m  + 3 \cdot N  \\+ (2N-1) \cdot L +L \cdot N+ N \cdot \lceil \log_2(N) \rceil. 
\end{multline} 

 Hence, using IR-HARQ results in a  memory overhead of $(L+1+\lceil \log_2(N)\rceil)\cdot N$ attributed to the LUT for storage of \pcfrozen values ($L \cdot N$), binary encoding of \pcfrozen bits ($N$) and LUT for storage of bit mappings ($\lceil \log_2(N)\rceil \cdot N$). We can see that the main contributor to memory overhead  is due to routing logic and simplifications should be done to reduce this. With $Q_i = 6, Q_e=5 , Q_m=8 , N=1024, L=8$, the memory overhead is $27\%$,while with $Q_i = 8, Q_e=5 , Q_m=11 , N=8192, L=8$ the memory overhead is around $25\%$.
\subsubsection{Computational Complexity}
\begin{table}[t!]
\caption{Computational complexity of HARQ operations per node}
\label{tab:comp-complexity}
\resizebox{\columnwidth}{!}{%
\bgroup
\def\arraystretch{1.5}%
\begin{tabular}{|l|l|}
\hline
\rowcolor[HTML]{EFEFEF} 
HARQ Node Operations      & SCL HARQ                                  \\ \hline
Ascending the \pcfrozen tree  & $4 \cdot  \frac{N_v}{2}  \cdot  \log (N_v) \cdot  C_{NAND}$       \\ \hline
Generating Candidates         & $4 \cdot  La \cdot L  \cdot C_{NAND}$                          \\ \hline
Descending the \pcfrozen tree   & $4 \cdot  \frac{N_v}{2}  \cdot  \log (N_v)  \cdot C_{NAND}$       \\ \hline
Routing \pcfrozen values                           & $    
 4\cdot \left(\sum_{i=sp}^{sp+2^s} 2^{\lceil(\log_2(i)\rceil} 
\cdot \lceil \log_2(i)\rceil\right)\cdot C_{NAND} $   \\ \hline
\end{tabular}%
\egroup
}
\end{table}




Finding out the exact computational complexity of the proposed HARQ scheme necessitates a hardware implementation since there can be a trade-off between latency, computational complexity, and throughput through the degree of parallelism used. For the current analysis, we assume a fully parallel implementation and we quantify complexity in terms of equivalent NAND gates. The following assumptions are used:
\begin{enumerate}
    \item Computation complexity of NOT, AND, OR, NOR and XOR gates are 1, 2, 3, 4, 4 $\times$ the computational complexity of a NAND gate ($C_{NAND}$) respectively. The latency of NOT, AND, OR, NOR and XOR gates are 1, 2, 2, 3, 3 $\times$ the latency of a NAND respectively.
    \item Computational complexity and latency of two input multiplexer (MUX) is 4 NAND gates and 3 NAND gates respectively. Larger input MUX can be built using a tree MUX structure requiring $2^{\lceil(\log_2(i)\rceil}\cdot \lceil \log_2(i)\rceil$ 2-input MUXes.
    \item Computational complexity of the comparison ($C_{comparison}$) of two 6-bit fixed point numbers in signed-magnitude notation is 45 NAND gates
    \item A bitonic sorter is used to sort surviving paths with a sorting computational complexity : \begin{multline}
     C_{sorting}=(\frac{La \cdot L}{4} \cdot \log_2(La \cdot L) \cdot (\log_2(La \cdot L)+1)\\ \cdot Q_{pm}) \cdot C_{comparison} .
     \label{eq:comp_bitonic}
\end{multline}
     \item The computational complexity of a half adder is $6\cdot C_{NAND}$. 

\end{enumerate}
 The proposed method adds two additional steps to the control flow. For each, the complexity overhead will be described below
 \begin{enumerate}
     \item Generating Bit Type:  The most complex operation in Section \ref{sec:generating-bit-types} is the addition operation  \ref{eq:pc_res}. A naive implementation of this function can be done by cascaded half adders where at each bit $ i>2$, $\lceil \log_2 (i+1) \rceil $  are needed. The computational complexity of a fully parallel implementation of that half adder structure is :
 \begin{equation}
     \begin{split}
     C_{Accumulator}=&\left[\left(\sum_{i=3}^{i=N} (\lceil \log_2 (i+1) \rceil) \cdot6 \right)+6 \right]\\
     &\cdot C_{NAND}.
     \end{split}
 \end{equation}


\item  Node Operations: The additional operations done with HARQ at each node alongside their computational complexities are mentioned in  \autoref{tab:comp-complexity}. Routing can be done using multiplexers; however, more computationally efficient memory manipulations can be done to circumvent the need of multiplexers at the cost of higher latency. 
 Encoding (ascending the tree) and Decoding (descending the tree) can be done sequentially through a $\log(Nv) \times \frac{Nv}{2}$ XOR tree while generating candidates can be done in parallel through the use of $La \times N_v$ XOR gates.

 \end{enumerate}

\subsubsection{Impact on latency}
The introduction of the linearity method is expected to increase latency as the list of candidate codewords at each node depends on the results of the previous node, and the computation of the candidates occurs along the critical path. 
The encoding and decoding of polar codes can be done using a XOR array with a latency of $4 \log(N_v) $ NAND gates. Generating candidates can be done in parallel with a latency of $4 \log(N_v) \cdot $ NAND gates. Routing can be done using multiplexers with a maximum latency of $3\cdot \lceil\log_2(N) \rceil$ NAND gates. 

\subsection{Performance Evaluation}
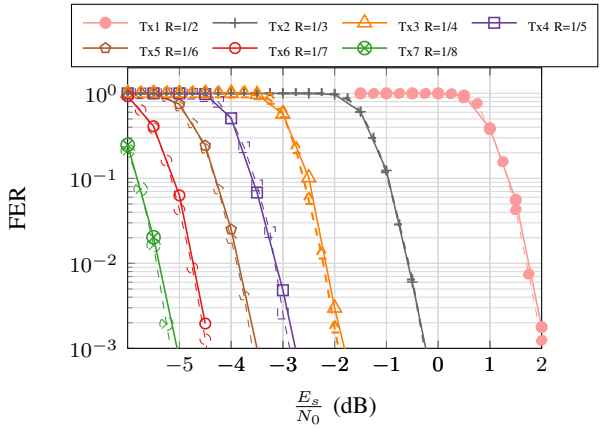
\begin{figure}[!t]
  \centering
  \begin{tikzpicture}[spy using outlines = {rectangle, magnification=2.0, connect spies}]
    \begin{groupplot}[group style={group name=fer_queries, group size= 2 by 1, horizontal sep=5pt, vertical sep=5pt},
      footnotesize,
      height=.6\columnwidth,  width=0.80\columnwidth,  
      xlabel=$\frac{E_s}{N_0}$ (dB),
      xmin=-6, xmax=2, xtick={-5,...,1,...,7},
      ymode=log,
      tick align=inside,
      grid=both, grid style={gray!30},
      /pgfplots/table/ignore chars={|},
      ] 

      \nextgroupplot[ylabel= FER, ytick pos=left, y label style={at={(axis description cs:-0.225,.5)},anchor=south},ymin=1e-3, ymax = 2]
      \addplot[mark=* , Paired-6, semithick]  table[x=Es/N0, y=FER] {plots/mine/Q2048.tex};\label{gp:p2}
      \addplot[mark=+, Paired-12, thick, semithick]  table[x=Es/N0, y=FER] {plots/mine/Q3072.tex};\label{gp:p3}
      \addplot[mark = triangle,mark options={scale=1.5}, Paired-7, thick, semithick]  table[x=Es/N0, y=FER] {plots/mine/Q4092.tex};\label{gp:p4}
      \addplot[mark=square,mark options={scale=1.0}  , Paired-9 , semithick]  table[x=Es/N0, y=FER] {plots/mine/Q5120.tex}; \label{gp:p5}
      \addplot[mark=pentagon, Paired-11, semithick]  table[x=Es/N0, y=FER] {plots/mine/Q6144.tex}      ;\label{gp:p6};
      \addplot[mark=o, Paired-5, semithick]  table[x=Es/N0, y=FER] {plots/mine/Q7168.tex} ;\label{gp:p7}
       \addplot[mark=otimes ,mark options={scale=1.25} , Paired-3, semithick]  table[x=Es/N0, y=FER] {plots/mine/Q8192.tex}  ;\label{gp:p8}
    \addplot[mark=* , Paired-6, dashed]  table[x=Es/N0, y=FER] {plots/theirs/2048.tex};
      \addplot[mark=+, Paired-12, thick, dashed]  table[x=Es/N0, y=FER] {plots/theirs/3072.tex};
      \addplot[mark = triangle,mark options={scale=1.5}, Paired-7, thick, dashed]  table[x=Es/N0, y=FER] {plots/theirs/4096.tex};
      \addplot[mark=square,mark options={scale=1.0}  , Paired-9 , dashed]  table[x=Es/N0, y=FER] {plots/theirs/5120.tex}; 
      \addplot[mark=pentagon, Paired-11, dashed]  table[x=Es/N0, y=FER] {plots/theirs/6144.tex}      ;
      \addplot[mark=o, Paired-5, dashed]  table[x=Es/N0, y=FER] {plots/theirs/7168.tex} ;
       \addplot[mark=otimes ,mark options={scale=1.25} , Paired-3, dashed]  table[x=Es/N0, y=FER] {plots/theirs/8192.tex}  ;

      \coordinate (top) at (rel axis cs:0,1);

      \coordinate (spypoint1) at (axis cs:7.45,2e-7);
      \coordinate (magnifyglass1) at (axis cs:2.6,1.1e-5);
      \coordinate (bot) at (rel axis cs:1,0);
    \end{groupplot}
    \path (top|-current bounding box.north) -- coordinate(legendpos) (bot|-current bounding box.north);
    \matrix[
    matrix of nodes,
    anchor=south,
    draw,
    inner sep=0.2em,
    draw
    ]at(legendpos) 
    {
    \ref{gp:p2}& \tiny  Tx1 R=1/2 &[1pt] 
    \ref{gp:p3}& \tiny  Tx2 R=1/3 &[1pt]
    \ref{gp:p4}& \tiny  Tx3 R=1/4 &[1pt]
    \ref{gp:p5}& \tiny  Tx4 R=1/5 \\  
    \ref{gp:p6}& \tiny  Tx5 R=1/6 &[1pt] 
    \ref{gp:p7}& \tiny  Tx6 R=1/7 &[1pt] 
    \ref{gp:p8} & \tiny Tx7 R=1/8 \\
      };
  \end{tikzpicture}
  \vspace*{-1em}
  \caption{\label{fig:FER_polar} Comparison of decoding performance of our proposed polar IR-HARQ scheme (straight lines) with the scheme in \cite{zhao_adaptive_2018} (dashed lines).}
   \vspace*{-1.3em}
\end{figure}

Fig. \ref{fig:FER_polar} shows the performance of our proposed method (solid lines) to the method in \cite{zhao_adaptive_2018} (dashed lines) in an AWGN channel with Quadrature Phase Shift Keying modulation. We use SCL $L=8$ and up to 7 transmissions (Tx)s, where the initial transmission is polar code (2048,1024+24) with initial rate (R) of $\frac{1}{2}$ constructed using Gaussian approximation \cite{trifonov_efficient_2012} and each re-transmission is of length 1024. Our simulation results use a quantization of $Q_e=5, Q_i=8$ and $ Q_m=11$ while the results of \cite{zhao_adaptive_2018} are in floating point non-quantized.  We can see that that the performance of both schemes is identical.

\section{Conclusion}
In this work, we present a series of hardware-friendly modifications to the polar IR-HARQ scheme \cite{zhao_adaptive_2018}. Our proposed changes, including a novel bit-type generation method and a modified control sequence for node processing, eliminates the reliance of \cite{zhao_adaptive_2018} on hardware-unfriendly set-based operations. Additionally the proposed candidate generation method allows the use of fast nodes even with the newly introduced bit-types. Our optimizations, resulting in a memory overhead of $27\%$ and $25\%$ for code lengths of $1024$ and $8192$ respectively, enhance the viability of polar IR-HARQ schemes for real-world deployment in next-generation communication systems such as 6G. Future research could focus on creating a hardware architecture employing our proposed techniques where we can accurately find the latency, area and throughput penalty of polar IR-HARQ.

\bibliographystyle{ieeetr}
\bibliography{IEEEabrv,new}

\begin{thebibliography}{10}

\bibitem{arikan_channel_2009}
E.~Arikan, ``Channel {Polarization}: {A} {Method} for {Constructing} {Capacity}-{Achieving} {Codes} for {Symmetric} {Binary}-{Input} {Memoryless} {Channels},'' {\em {IEEE} Trans. Inf. Theory}, vol.~55, pp.~3051--3073, July 2009.

\bibitem{tal_list_2015}
I.~Tal and A.~Vardy, ``List {Decoding} of {Polar} {Codes},'' {\em {IEEE} Trans. Inf. Theory}, vol.~61, no.~5, pp.~2213--2226, 2015.

\bibitem{3gpp_nr_2021}
3GPP, ``{NR} {Multiplexing} and channel coding,'' Technical {Specification} ({TS}) 3GPP.38.212, 3rd Generation Partnership Project, Nov. 2021.

\bibitem{cheng_coding_2006}
J.-F. Cheng, ``Coding performance of hybrid {ARQ} schemes,'' {\em {IEEE} Trans. Commun.}, vol.~54, pp.~1017--1029, June 2006.

\bibitem{chen_hybrid_2013}
K.~Chen, K.~Niu, and J.~Lin, ``A {Hybrid} {ARQ} {Scheme} {Based} on {Polar} {Codes},'' {\em {IEEE} Commun. Lett.}, vol.~17, pp.~1996--1999, Oct. 2013.

\bibitem{chen_polar_2014}
K.~Chen, K.~Niu, Z.~He, and J.~Lin, ``Polar coded {HARQ} scheme with {Chase} combining,'' in {\em {IEEE} {Wireless} {Commun.} and {Netw.} {Conf.}}, pp.~474--479, Apr. 2014.

\bibitem{el-khamy_harq_2015}
M.~El-Khamy, H.-P. Lin, J.~Lee, H.~Mahdavifar, and I.~Kang, ``{HARQ} {Rate}-{Compatible} {Polar} {Codes} for {Wireless} {Channels},'' in {\em {IEEE} {Global} {Commun.} {Conf.}}, pp.~1--6, Dec. 2015.

\bibitem{tavildar_h-arq_2016}
S.~R. Tavildar, ``A {H}-{ARQ} scheme for polar codes,'' June 2016.
\newblock arXiv:1606.08545.

\bibitem{zhang_hybrid_2018}
Y.~Zhang, K.~Qin, C.~Jiao, and Z.~Zhang, ``A {Hybrid} {ARQ} {Scheme} {Based} on {Equivalent} {Puncturing} {Patterns} of {Polar} {Codes},'' in {\em {IEEE} Commun. Lett.}, pp.~1--5, Aug. 2018.
\newblock ISSN: 2577-2465.

\bibitem{chen2022equivalent}
K.~Chen, C.~Xu, W.~Liangming, J.~Li, and H.~Xu, ``Equivalent puncture sets for polar coded re-transmissions,'' Aug.~9 2022.
\newblock US Patent 11,411,678.

\bibitem{zhao_adaptive_2018}
M.-M. Zhao, G.~Zhang, C.~Xu, H.~Zhang, R.~Li, and J.~Wang, ``An {Adaptive} {IR}-{HARQ} {Scheme} for {Polar} {Codes} by {Polarizing} {Matrix} {Extension},'' {\em {IEEE} Commun. Lett.}, vol.~22, pp.~1306--1309, July 2018.

\bibitem{li_capacity-achieving_2016}
B.~Li, D.~Tse, K.~Chen, and H.~Shen, ``Capacity-achieving rateless polar codes,'' in {\em {IEEE} {International} {Symposium} {Inf.} {Theory}}, pp.~46--50, July 2016.

\bibitem{leroux_semi-parallel_2013}
C.~Leroux, A.~J. Raymond, G.~Sarkis, and W.~J. Gross, ``A {Semi}-{Parallel} {Successive}-{Cancellation} {Decoder} for {Polar} {Codes},'' {\em {IEEE} Trans. Signal Process.}, vol.~61, pp.~289--299, Jan. 2013.

\bibitem{sarkis_fast_2014}
G.~Sarkis, P.~Giard, A.~Vardy, C.~Thibeault, and W.~J. Gross, ``Fast polar decoders: Algorithm and implementation,'' vol.~32, no.~5, pp.~946--957.

\bibitem{balatsoukas-stimming_hardware_2014}
A.~Balatsoukas-Stimming, A.~J. Raymond, W.~J. Gross, and A.~Burg, ``Hardware architecture for list successive cancellation decoding of polar codes,'' {\em {IEEE} Trans. Circuits Syst. {II}}, vol.~61, pp.~609--613, Aug. 2014.

\bibitem{sarkis_fast_2016}
G.~Sarkis, P.~Giard, A.~Vardy, C.~Thibeault, and W.~J. Gross, ``Fast {List} {Decoders} for {Polar} {Codes},'' {\em {IEEE} J. Sel. Areas Commun.}, vol.~34, pp.~318--328, Feb. 2016.

\bibitem{hashemi_fast_2017}
S.~A. Hashemi, C.~Condo, and W.~J. Gross, ``Fast and {Flexible} {Successive}-{Cancellation} {List} {Decoders} for {Polar} {Codes},'' {\em {IEEE} Trans. Signal Process.}, vol.~65, pp.~5756--5769, Nov. 2017.

\bibitem{hashemi_fast_2019}
S.~Hashemi, {\em Fast, {Flexible}, and {Area}-efficient {Decoders} for {Polar} {Codes}}.
\newblock {McGill} theses, McGill University Libraries, 2019.

\bibitem{trifonov_efficient_2012}
P.~Trifonov, ``Efficient {Design} and {Decoding} of {Polar} {Codes},'' {\em {IEEE} Trans. Commun.}, vol.~60, pp.~3221--3227, Nov. 2012.

\end{thebibliography}
\end{document}